\documentstyle[12pt]{article}
\input{epsf}
\begin{document}

\def\fa{Fig.~1(a)} \def\fb{Fig.~1(b)}
\def\lg{{\langle }}
\def\mt{\mapsto }
\def\od{\odot }
\def\ra{{\rightarrow }}
\def\rap{\ \rightarrow\ }
\def\Ra{{\Rightarrow }}
\def\rg{{\rangle }}
\def\srt{\sqrt{2}}
\def\vb{{\,|\,}}
\def\Tr{\hbox{Tr}}

\title{Consistent Histories and Quantum Delayed Choice}

\author{Robert B. Griffiths\thanks{Electronic mail: rgrif@cmu.edu}\\
Department of Physics,
Carnegie-Mellon University,\\
Pittsburgh, PA 15213}

\date{To appear in Fortschritte der Physik}

\maketitle

\begin{abstract}
		John Wheeler devised a gedanken experiment in which a piece of
apparatus can be altered just before the arrival of particle, and this ``delayed
choice'' can, seemingly, alter the quantum state of the particle at a much
earlier time, long before the choice is made.
	A slightly different gedanken experiment, which exhibits the same
conceptual difficulty, is analyzed using the techniques of consistent history
quantum theory.  The idea that the future influences the past disappears when
proper account is taken of the diversity of possible quantum descriptions of
the world, and their mutual compatibility or incompatibility. 
\end{abstract}

	John Wheeler proposed a delayed-choice experiment in which the second
beam splitter of a Mach-Zehnder interferometer is or is not removed at the very
last instant before a photon reaches it and passes out of the interferometer
\cite{whe}.  A modified version which exhibits essentially the same paradox is
shown in Fig.~1.  A photon enters the interferometer through channel $a$, and
two detectors in the $c$ and $d$ arms are either (a) left in position, or (b)
removed just before the photon reaches (one of) them, so that it is free to
proceed on to the second beam splitter.  To be sure, physically moving
detectors out of the way at the last instant is a theorist's fantasy, but
clever experimentalists can do essentially the same thing using Pockels cells
and polarizers \cite{all,hw87}.

	The paradox arises in the following way.  Suppose detectors $C$ and $D$
are present, \fa, and one of them, say $C$, detects the photon.  Then it seems
plausible that at an earlier time the photon was in the $c$ arm of the
interferometer. There are various ways in which this idea could be supported
experimentally.  For example, by seeing what happens when one removes mirror
$M_c$, or mirror $M_d$, or inserts an absorber in the $c$ or $d$ arm of the
interferometer upstream from the detectors, or lengthens the path in one arm
while timing when the photon reaches the detector.  These considerations make
it plausible that if one of the detectors detects the photon, the latter was,
at an earlier time, in the corresponding arm of the interferometer.


	On the other hand, if $C$ and $D$ are out of the way, \fb, removed at
the very last instant, the second beamsplitter and detectors $E$ and $F$ which
follow it can be used to test whether the photon was in a coherent
superposition state between the two arms while inside the interferometer.  To
be specific, suppose the action of the first beam splitter is described by
the unitary transformation
\begin{equation}
  |a\rg \mt |s\rg = (|c\rg + |d\rg )/\srt,
\label{e1}
\end{equation}  
where $|a\rg$ is a wave packet in the entrance channel of the interferometer,
and $|c\rg$ and $|d\rg$ are wave packets in the $c$ and $d$ arms, respectively.
Similarly, assume a unitary transformation
\begin{equation}
  |c\rg \mt |u\rg = {(|e\rg + |f\rg)/  \srt}, \quad
  |d\rg \mt |v\rg = {(-|e\rg + |f\rg)/ \srt}
\label{e2}
\end{equation}
at the second beamsplitter.
As a consequence, 
\begin{equation}
  |a\rg\mt |s\rg \mt |f\rg,
\label{e3}
\end{equation}
so that a photon which enters the interferometer in channel $a$ will emerge
in channel $f$, to be detected by detector $F$ and not by detector $E$.  If the
experiment is repeated many times, and if every time $C$ and $D$ are pushed out
of the way the photon is later detected by $F$ rather than $E$, then it is
plausible that inside the interferometer the photon was in the coherent
superposition state $|s\rg$ with a definite phase relationship between the wave
packets in the two arms, rather than entirely in the $c$ arm or in the $d$ arm.
Indeed, it is hard to see how else one could explain the outcome.

	But now we have a paradox.  For it the detectors $C$ and $D$ are left
in place, the photon was (plausibly) in either the $c$ or the $d$ arm, that is,
in the arm in which it will later be detected, ever since it left the first
beamsplitter.  On the other hand, if $C$ and $D$ are removed at the very last
moment, then, in order that it be later detected by $F$ and not by $E$, the
photon needs to have been in the superposition $|s\rg$, and this must have been
the case ever since it left the first beam splitter.  Consequently, the photon
when it passed the first beamsplitter had to ``know'' whether the detectors $C$
and $D$ would or would not be present later on.

	Wheeler used this paradox to argue, in effect, that the process by
which the photon moves through the interferometer cannot be described or
thought about in a coherent way.  It is, to use his words, a ``great smoky
dragon'': one cannot say anything at all about what is going on before the
dragon ``bites'' one of the detectors, thus bringing the experiment to a close
by an irreversible act of measurement.  Or, if I may use a different metaphor,
a quantum system is like a ``black box'', and one should not try and figure out
what is going on inside the box.  

	The whole field of quantum foundations is littered with these black
boxes, and physicists foolish enough to open them have often been stung by the
paradoxes (or burned by the dragons) which then emerge.  Although there are
important differences, Wheeler's paradox, in which the future seems to
influence the past, bears a certain resemblance to the Elitzur-Vaidman paradox
of ``interaction free measurement'' \cite{ev93}, a delayed-choice version of
which is obtained by discarding one of the counters (say $D$) in \fa, and
Hardy's paradox \cite{hd92}, in which there seems to be a mysterious long-range
influence in violation of relativity theory.

	The consistent history formalism 
\cite{gr84,gh90,om92,om94}
is ideally suited to addressing problems of this sort, because it allows one to
discuss what is going on in a closed quantum mechanical system in a realistic
way, without running into logical inconsistencies.  One can, so-to-speak, open
the black box and disarm the paradox without being stung (or burned).  Omn\`es'
book \cite{om94b} discusses a modified form of Wheeler's paradox from the
consistent history perspective.  While it is technically sound (once some
misprints have been corrected \cite{om94c}), this treatment is not entirely
satisfactory, because it relies on a distinction between ``true'' and
``reliable'' which both Omn\`es \cite{ompc} and his critics \cite{dk956,gr95}
agree is flawed.  Consequently, it is worthwhile exploring Wheeler's paradox
once again in the light of a recent restatement of the principles of consistent
history reasoning found in \cite{gr96}, and augmented in \cite{gr97}, which (I
believe) does not suffer from the problem just referred to.  (Incidentally,
\cite{gr97} is also a response to certain criticisms of the consistent history
approach found in \cite{dk956}.)

	As usual, when employing consistent histories the entire
system, including detectors, must be treated quantum mechanically.  Thus for
the situation in Fig.~1, let $|C\rg$ indicate the ``ready'' state of detector
$C$ before a photon has arrived, $|C^*\rg$ the corresponding state in which it
has detected a photon, and let the unitary transformation representing
detection be given by
\footnote{More realistic models of measurement can be
employed in the consistent history formalism; for an example, see Sec.~VI~C of
\cite{gr96}.}
\begin{equation}
  |c\rg|C\rg\mt |C^*\rg.
\label{e4}
\end{equation}
A similar notation is employed for the other detectors.  One should think of
$|C\rg$ and $|C^*\rg$ as macroscopically distinct states; e.g., in the quaint
but picturesque language of quantum foundations, there is a very visible
pointer attached to the detector, which (say) points down for $|C\rg$ and up
for $|C^*\rg$.

	A consistent history analysis of the situation in \fa\ proceeds by
introducing a {\it consistent family} of possible histories. There are many
different ways to choose a family; here is one of them:
\begin{equation}
  |a\rg|CD\rg\rap\cases{ |c\rg|CD\rg\rap |C^*D\rg, &\cr
		|d\rg|CD\rg\rap |CD^*\rg, & \cr}
\label{e5}
\end{equation}
where the notation is to be interpreted in the following way.  Time progresses
from left to right, starting at $t_0$ when the photon is in a wave packet
$|a\rg$, that is, about to enter the interferometer, and detectors $C$ and $D$
are ready.  At the next time, $t_1$, the photon is either in a wave packet
$|c\rg$ in the $c$ arm, or else a wave packet $|d\rg$ in the $d$ arm.  If it is
in $|c\rg$ at $t_1$, then at a later time $t_2$ it will have triggered the $C$
detector, changing it from $|C\rg$ to $|C^*\rg$, while detector $D$ remains in
its untriggered (``ready'') state.  Similarly, in the second history in
(\ref{e5}), the photon is in $|d\rg$ at $t_1$, and has triggered $D$, and not
$C$, at time $t_2$.  The representation (\ref{e5}) is somewhat abbreviated in
that the beam splitters and the detectors $E$ and $F$, which play only a
passive role, are not shown explicitly.  If we put all of these into a single
initial state $|\Psi\rg$, then, in the notation of \cite{gr96}, family
(\ref{e5}) contains the two histories,
\begin{equation}
  \Psi\od c\,CD\od C^*D,\quad \Psi\od d\,CD\od CD^*.
\label{e6}
\end{equation}
But we will not use (\ref{e6}) in the subsequent discussion, so the reader
unfamiliar with this notation can ignore it and simply refer to (\ref{e5}).

	It is important to notice that the two histories in (\ref{e5}), even
though they start with the same initial state, are {\it mutually exclusive}
alternatives (a point which is, perhaps, a bit clearer in the notation of
(\ref{e6})): if the photon is in the $c$ arm at time $t_1$, it is definitely
not in the $d$ arm, and it will later trigger detector $C$, and not trigger
detector $D$. By applying an appropriate probability calculus, as explained in
\cite{gr96}, to (\ref{e5}), one can conclude that in the case in which the
photon was detected by detector $C$, it definitely (that is, with conditional
probability equal to one) was in the $c$ arm at an earlier time, just as a
naive physicist would tend to believe (at least when not being chased by a
smoky dragon!).  Thus in this particular instance, consistent histories
supports the usual intuition employed for designing experimental apparatus.  We
have opened the black box, and nothing has gone wrong---at least, not yet.

	But how are we going to handle the situation in \fb?  Again, it is
necessary to choose a consistent family, and one which will work very well for
this purpose is:
\begin{equation}
    |a\rg|EF\rg\rap |s\rg|EF\rg\rap 
	\cases{ |e\rg|EF\rg\rap |E^*F\rg, &\cr
		|f\rg|EF\rg\rap |EF^*\rg, & \cr}
\label{e7}
\end{equation}
where there are now four different times: the initial time $t_0$, $t_1$ when
the photon is in the superposition state $|s\rg$ (defined in (\ref{e1})) inside
the interferometer, $t_2$ when it has passed through the second beam splitter
to emerge in either the $e$ or $f$ channel, and $t_3$ when it has been detected
by either $E$ or $F$.  In fact, with the unitary transformations indicated in
(\ref{e1}) and (\ref{e2}), the photon will surely emerge in $|f\rg$, see
(\ref{e3}).  As a consequence, the upper history in (\ref{e7}) has probability
0, so it will never occur, while the lower history, ending with $F^*$, will
occur with probability 1.  (Actually, we omitted from (\ref{e5}) a certain
number of histories whose presence is needed for a consistent formalism
according to the rules of
\cite{gr96}, but which occur with probability zero, because they are dynamically
impossible, so we could also have omitted the upper history in (\ref{e7}).)
By considering the family (\ref{e7}) we can conclude from the fact that the
photon was detected by $F$, or simply on the basis of the initial state, that
the photon was certainly (probability 1) in the coherent superposition $|s\rg$
while inside the interferometer.  So once again we have opened up the black
box, and found a physically reasonable result.

	But something looks suspicious.  For the case in which detectors $C$
and $D$ remain in place, we used (\ref{e5}), in which the photon is definitely
in the $c$ or $d$ arm after it passes through the first beam splitter, whereas
for the situation in which both $C$ and $D$ are removed at the last moment, we
used (\ref{e7}), in which the photon is in a coherent superposition all the
time it is inside the interferometer.  Isn't this precisely Wheeler's paradox?
No, because the choice of family is one made by the physicist in constructing a
(possible) description of the quantum system; it is not something forced upon
him either by a law of nature or by later events in the system under
consideration.  To see that this is so, let us examine what happens if, in
place of (\ref{e7}), we use a consistent family of histories in which the
photon is in a definite arm, $c$ or $d$, while it is inside the interferometer:
\begin{equation}
    |a\rg|EF\rg\rap 
	\cases{ |c\rg|EF\rg\rap |u\rg|EF\rg\rap |U\rg, &\cr
		|d\rg|EF\rg\rap |v\rg|EF\rg\rap |V\rg. & \cr}
\label{e8}
\end{equation}
Here the four successive times are the same as in (\ref{e7}), $|u\rg$ and
$|v\rg$ were defined in (\ref{e2}), and
\begin{equation}
 |U\rg = {(|E^*F\rg + |EF^*\rg)/ \srt}, \quad
  |V\rg = {(-|E^*F\rg + |EF^*\rg)/ \srt}
\label{e9}
\end{equation}	
are macroscopic quantum superposition (MQS) or Schr\"odinger cat states, since
they are coherent superpositions of macroscopically distinct situations. 

	Both families (\ref{e7}) and (\ref{e8}) apply to the same dynamical
situation (detectors $C$ and $D$ out of the way, \fb), and from the point of
view of consistent histories, either is equally valid as a description of the
quantum world.  To be sure, (\ref{e8}) involves MQS states at the final time
$t_3$, and we shall say more about this later.  However, it is just as
significant that at time $t_1$, that is, while the photon is inside the
interferometer, (\ref{e7}) assigns to it a state $|s\rg$, and (\ref{e8}) one of
the two states $|c\rg$ or $|d\rg$.  Surely these cannot both be right. Is it
(\ref{e7}) or is it (\ref{e8}) which tells us what the photon is {\it really}
doing inside the interferometer?

	The question just posed {\it has no answer}, and it is important to
understand why that is so, as it goes to the very heart of what distinguishes
quantum theory from classical physics.  Consider the simplest of all quantum
systems, a spin half particle with a two-dimensional Hilbert space, where each
ray (i.e., one-dimensional subspace) corresponds to a spin angular momentum of
+1/2 (in units of $\hbar$) in a particular direction.  In standard
quantum mechanics (no hidden variables) the statement ``$S_x=1/2$'' makes
sense, for it corresponds to something definite in the Hilbert space, a
particular ray (i.e., a particular ket, up to multiplication by an arbitrary
complex number), and so, of course, does ``$S_z=1/2$''.  On the other hand,
``$S_x=1/2$ AND $S_z=1/2$'' does not make sense, for there is no ray which
corresponds to it.%
\footnote{Since every ray corresponds to the spin
being in a particular direction, there are none left over to represent two
different directions joined by AND. For a fuller treatment of
this point, see \cite{gr97}, Sec.~IV~A} 
But if connecting two propositions with AND makes no sense, connecting them
with OR is no better---at least this is the consistent histories
perspective---and thus the question ``is the spin in the $+z$ direction or is
it in the $+x$ direction?'' is meaningless (in the sense that quantum theory
ascribes it no meaning).  And, by analogy, the question ``Was the photon in
$|s\rg$ or in $|c\rg$ at $t_1$?'' makes no sense.  In the technical terminology
of consistent histories, the families (\ref{e7}) and (\ref{e8}) are {\it
incompatible\/}: descriptions based upon one cannot be combined with those
based upon the other, nor does it make sense to ask ``which is right?''.  For
further details, see \cite{gr97}.

	To be sure, Wheeler's ``great smoky dragon'' is also a way of saying
that certain things cannot be discussed in quantum theory, certain things do
not make sense.  The difference is that in the consistent history approach the
rules as to what makes sense and what does not are precise, and based upon the
mathematical structure of quantum theory itself.  They do not take the form of
a blanket prohibition: ``Don't talk about microscopic systems, because that is
dangerous.''  Instead, they are of the form: ``This makes sense (if you use a
Hilbert space) but that doesn't make sense,'' or: ``This family satisfies the
consistency conditions, so probabilities can be assigned, whereas that other
family does not satisfy the consistency conditions.''  The basic point is that
if one employs, as in standard quantum mechanics, a Hilbert space and unitary
time transformations to produce descriptions of a quantum system, the logical
rules for interpreting these descriptions should be compatible with the
underlying mathematics.  Failure to pay attention to this requirement is the
source of many paradoxes, superluminal influences, and other quantum ghosts.

	We have seen that the absence of detectors $C$ and $D$ does not force
one to adopt a coherent superposition state $|s\rg$ for the photon while it is
inside the interferometer; (\ref{e8}) can be used rather than (\ref{e7}).
Similarly, when $C$ and $D$ are present, one does not have to employ a family,
such as (\ref{e5}), in which the photon is in one arm or the other.  Here is a
perfectly good alternative:
\begin{equation}
  |a\rg|CD\rg\rap |s\rg|CD\rg \rap |S\rg,
\label{e10}
\end{equation}
where 
\begin{equation}
  |S\rg = (|C^*D\rg + |CD^*\rg)/\srt
\label{e11}
\end{equation}
is another MQS state; as before, various histories with zero probability
belonging to this family are not shown explicitly.  Once again, there is no
``law of nature'' which singles out  (\ref{e5}) in contrast to  (\ref{e10}) as
the  ``correct'' consistent family; from the perspective of fundamental quantum
theory, both are equally correct, both provide perfectly good descriptions of
the quantum world.  However, they are incompatible (remember $S_x$ and $S_z$!),
so that only one, not both, can be used to describe a given situation. 

	But if the consistent histories approach allows many different
consistent families, and gives no procedure for specifying which is the right
one, how can it be a rational scientific theory of the world?  Examining the
different consistent families introduced above, one sees that certain families
can address certain questions, and other families can address other questions.
Suppose, for example, one wants to know whether it is detector $E$ or detector
$F$ which detects the photon in \fb.  While (\ref{e7}) and (\ref{e8}) are
equally good consistent families from the perspective of fundamental quantum
theory, the question of ``which detector'' can be addressed using (\ref{e7}),
but not (\ref{e8}). The reason is that the MQS states $|U\rg$ and $|V\rg$ in
(\ref{e9}) are incompatible (in the quantum sense) with a discussion of whether
detector $E$ has or has not detected a photon; think of $|U\rg$ as something
like $S_x=1/2$ and $|EF^*\rg$ as something like $S_z=1/2$.  Similarly, if we
are interested in whether $C$ or $D$ detected the particle in \fa, we can use
(\ref{e5}), but we cannot use (\ref{e10}).  However, (\ref{e5}) is not the only
possibility for discussing ``$C$ or $D$?''; the consistent family
\begin{equation}
    |a\rg|CD\rg\rap |s\rg|CD\rg\rap 
	\cases{ |C^*D\rg, &\cr
		 |CD^*\rg, & \cr}
\label{e12}
\end{equation}
will work equally well.  An important difference between (\ref{e5}) and
(\ref{e12}), however, is that the former allows us to address the question ``in
which arm of the detector was the photon before it was detected?'', whereas the
latter does not: $|c\rg$ and $|d\rg$ are incompatible with $|s\rg$ at the time
$t_1$.

	Consequently, one sees that the multiplicity of quantum descriptions,
that is, consistent families, allowed by the consistent history approach is
sharply narrowed as soon as we focus on particular questions of physical
interest.  Some families can be used to address certain questions, other
families are needed to address other questions.  In addition,
when more than one consistent family which can be used to address a
particular physical question---e.g., (\ref{e5}) and (\ref{e12}) can both be
used to answer ``$C$ or $D$?''---the answers (in terms of probabilities) given
by the different families are the same; see Sec.~IV of \cite{gr96}.

	The various consistent families introduced above show that opening the
black box of quantum theory does {\it not} lead to the conclusion that the
future influences the past, for we can adopt either a coherent superposition
description, $|s\rg$, or the one-arm-or-the-other description using $|c\rg$ and
$|d\rg$ for the photon inside the interferometer, whether or not the detectors
$C$ and $D$ will later be present. However, the presence or absence of the $C$
and $D$ detectors does have an important dynamical effect: compare
(\ref{e5}) with (\ref{e8}).%
\footnote{In order to keep the notation from becoming
unwieldy, we have omitted detectors $E$ and $F$ from (\ref{e5}), and $C$ and
$D$ from (\ref{e8}); putting them in causes no problem, and makes the
expressions look more similar.}
In particular, MQS states, $|U\rg$ and $|V\rg$,
appear in (\ref{e8}) but not in (\ref{e5}).  Physicists tend to find MQS states
embarrassing, and one can sympathize with a certain reluctance to use
(\ref{e8}), and a corresponding preference for (\ref{e7}).  However, this is
not a matter of ``future influencing past'', at least as a physical effect.  It
is more like what happens when an author is writing a novel and adjusts certain
events near the beginning so that the last chapter turns out the way he wants.
Note that the consistent history approach does not ``rule out'' MQS states;
indeed, they have to be present in the Hilbert space of standard quantum theory
precisely because it is a linear vector space.  However, they do not occur in
every consistent family, and for this reason the consistent history approach is
not troubled by the infamous ``measurement problem'' which has given rise to a
great deal of work, and an enormous amount of confusion, in the field of
quantum foundations; see the valuable critique by John Bell in \cite{bl90}.

	The analysis given above is incomplete in one respect: we have treated
the situations in Fig.~1(a) and (b) separately, and have not tried to put the
whole story together by, for example, introducing a quantum coin and an
associated servo mechanism which pushes $C$ and $D$ out of the way at the last
instant if the coin turns up heads, or leaves it in place if the result is
tails.  (While this may sound like another theorist's fantasy, the
experimentalists already know how to do what is essentially the same thing
\cite{all}.)  Putting the whole story together does not yield any new insights
into the question of whether the future influences the past, which I believe
can be answered (in the negative) on the basis of the considerations given
above.  However, it does open up additional possibilities for consistent
families, including the (analog of the) one discussed by Omn\`es \cite{om94d},
and allows one to pose counterfactual questions such as the following: Suppose
that $C$ and $D$ were left in place.  What {\it would have} happened {\it if}
the quantum coin had turned up the other way, and $C$ and $D$ had been removed
before the photon arrived?  Discussing this interesting question goes beyond
the scope of the present paper.

	In conclusion, the consistent histories approach provides an analysis
of the situation in Fig.~1 in terms of what the photon was doing inside the
interferometer, without running into a paradox in which the future influences
the past.  The key idea is to use consistent histories in order to have a
logical structure for quantum theory which is consistent with the mathematics,
and which can be used to discuss the time development of a quantum system
without running into paradoxes. 

	It is a pleasure to thank C. Alley for providing references to
experimental studies of delayed choice.  Financial support for this
research was provided by the National Science Foundation through grant PHY
96-02084.

\begin{figure}[h]
\epsfxsize=14truecm
\epsfbox{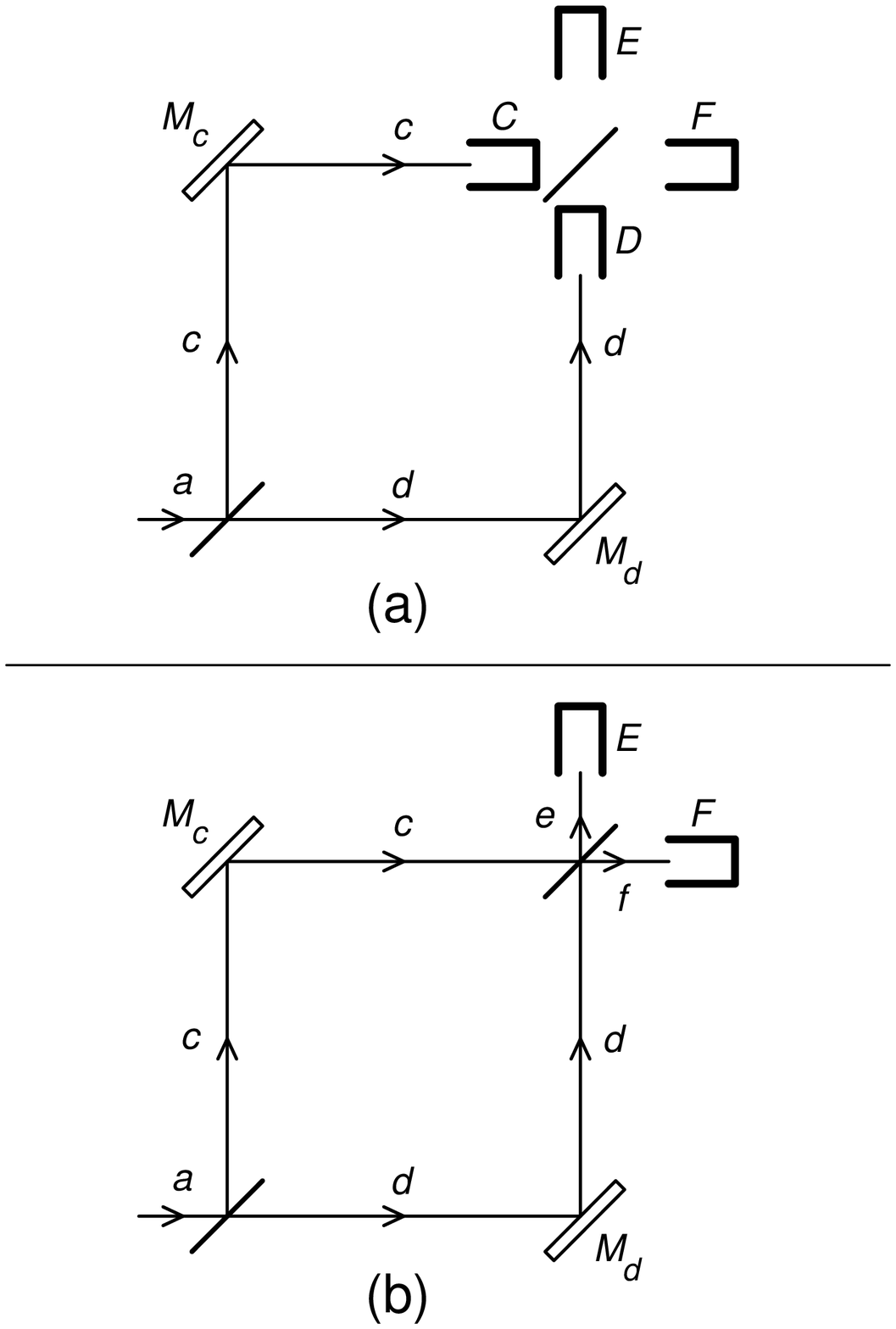}
\caption{Mach-Zehnder interferometer with (a) detectors in arms $c$ and $d$;
(b) detectors removed at the last moment.}
\end{figure}


\begin{thebibliography}{11}
\bibitem{om94c}
On the lower half of p. 462 of \cite{om94} replace $V_1$ and $V_2$ with $R_1$
and $R_2$, and note that the histories referred to as (a), (b), and (c) are
those numbered 1, 2, and 3, respectively, on the upper half of the page.
\end{thebibliography}
\end{document}